\documentclass[preprint,showpacs,preprintnumbers,amsmath,amssymb]{revtex4}

\usepackage{dcolumn}
\usepackage{bm}

\begin{document}


\title{Solutions To The Classical Liouville Equation}

\author{Jose A. Magpantay}
\email{jose.magpantay@upd.edu.ph}
\affiliation{National Institute of Physics and Science and Society Program, University of the Philippines, Diliman, Quezon City, Philippines\\}
\author{Cilicia Uzziel M. Perez}
\email{uzzie.perez@themindmuseum.org}
\affiliation{The Mind Museum, Bonifacio Global City and National Institute of Physics, University of the Philippines, Philippines\\}

%

%

\date{\today}

\begin{abstract}
We present solutions to the classical Liouville equation for ergodic and completely integrable systems - systems that are known to attain equilibrium. Ergodic systems are known to thermal equilibrate with a Maxwell-Boltzmann distribution and we show a simple derivation of this distribution that also leads to a derivation of the distribution at any time t. For illustrative purposes, we apply the methods derived to the problem of a one-dimensional gravitational gas even though its ergodicity is debatable. For completely integrable systems, the Liouville equation in the original phase space is rather involved because of the group structure of the integral invariants, which hints of a 'gauge' symmetry. We use Dirac's constrained formalism to show the change in the Liouville equation, which necessitates the introduction of gauge-fixing conditions. We then show that the solution of the Liouville equation is independent of the choice of gauge, which it must be because physical quantities are derived from the distribution. Instead, we derive the solution to the classical Liouville equation in the phase space where the dynamics involve ignorable coordinates, a technique that is akin to the use of the unitarity gauge in spontaneously broken theories to expose the physical degrees of freedom. It turns out the distribution is time-independent and precisely given by the generalized Gibbs ensemble (GGE), which was solved by Jaynes using the method of constrained optimization. As an example, we apply the method to the problem of two particles in 3D interacting via a central potential, which is a completely integrable system.     
\end{abstract}

\maketitle

\section{\label{sec:level1}Introduction}
The classical statistical mechanics of many particle systems begins with the Liouville equation, which solves for the distribution function in phase space \cite{Huang}, \cite{Reichl}. Since the equation is first order in time, we need the initial distribution of the system to solve for the distribution at any time t. The complications in solving this equation are two fold - the equation is a first order PDE with non-constant coefficients in general and the initial condition, because of the physics, is sometimes not classical \cite{Horenko}. We will not consider the case of quantum initial condition in this paper. The first complication leads to not having closed form solutions and use of approximation methods, for example, the gausssian phase packet dynamics \cite{Ma}.

	But before we even begin solving the Liouville equation, we have to establish the nature of the system's classical dynamics for it will determine if the distribution will even attain equilibrium. And as will be shown in this paper, the attainment of equilibrium suggests ways of solving the Liouville equation. 
	
	Classical systems are either integrable or non-integrable. The classification is determined by the existence of integral invariants, which determines the region of phase space accessed by the system in its evolution. Integrable systems that have the number of integral invariants equal to the number of degrees of freedom and the integral invariants satisfying an involution (defined below) are said to be a completely integrable. Non-integrable systems with chaotic dynamics are said to be ergodic. These two are the extreme behaviors a system can have. And it is these two extremes that have been shown to attain equilibrium. 
	
	Consider a system of N particles in one dimension. The phase space with coordinates and momenta $(x_i,p_i)$ with $i=1,..N$, is 2N dimensional. The system dynamics is defined by the Hamiltonian $H(x,p)$ with the evolution given by Hamilton's equations
\begin{subequations}\label{sha1}
\begin{gather}
\dot{x}_i=\dfrac{\partial H}{\partial p_i},\label{first}\\
\dot{p}_i=-\dfrac{\partial H}{\partial x_i}.
\end{gather}
\end{subequations}
For the system to be completely integrable, there must exist N integral invariants $I_i(x,p)$, which define constants of motion. Furthermore, the integral invariants must be in involution, i.e., the Poisson brackets vanish, see for example \cite{Masoliver}. 
\begin{equation}\label{sha2}
\left\{I_i,I_j\right\}=\sum_{k=1}^N\left\{\dfrac{\partial I_i}{\partial x_k}\dfrac{\partial I_j}{\partial p_k}-\dfrac{\partial I_i}{\partial p_k}\dfrac{\partial I_j}{\partial x_k}\right\}=0.
\end{equation} 
The integral invariants define N tori \cite{Arnold} in phase space and these represent where the system can be found as the particles evolve according to equation (1). Since the system covers only a limited region of the phase space defined by the invariant tori, it can only equilibrate as described by the distribution given by the generalized Gibbs ensemble
\begin{subequations}\label{sha3}
\begin{gather}
f_{GGE}=Z^{-1}(a_i)\exp{\left(-\sum a_iI_i\right)},\label{first}\\
Z(a_i)=\int d^Nxd^Np\exp{\left(-\sum a_iI_i\right)}.
\end{gather}
\end{subequations}
The Lagrange multipliers $a_i$ are solved from the initial conditions through
\begin{equation}\label{sha4}
I_{0i}=\int d^Nxd^Np I_i(x,p)f_{GGE}.
\end{equation}
This equilibrium distribution can be derived with the method of optimization with constraints \cite{Jaynes}. 

	In this paper, we will present a direct derivation of the generalized Gibbs ensemble in the phase space with ignorable coordinates. We will also derive the canonical transformation properties of the distribution using the Liouville equation and the Jacobian of the phase space integral measure under canonical transformation. Using these results, we derive the generalized Gibbs ensemble.
	
	If a small perturbation to an integrable system is added, the KAM theorem says that most systems will evolve close to the region defined by the invariant tori, the orbit is stable. This also means the distribution is not exactly given by the GGE equilibrium but since the dynamics is not chaotic, the system cannot thermal equilibrate either. The big deviation off the invariant tori happens for resonant systems (the perturbation makes the system move far from the invariant tori surfaces) and if the number of degrees of freedom is at least two (phase space is at least four), providing an extra dimension off the invariant tori, the system can wander off to the other surfaces defined by the energy hypersurface (Arnold diffusion).   
	
	The extreme case of non-integrability is ergodicity, where the system covers the entire phase space, or the phase space defined by the conserved energy torus, because of chaotic dynamics. In this case, the system attains thermal equilibrium with a Maxwell-Boltzmann distribution $f \propto \exp{\left(-\beta H\right)}$. We present a general solution of the distribution valid at any time, which also clearly shows the equilibrium distribution. 
	
	The question then is given a Hamiltonian, how do we tell whether the system is integrable, non-integrable or ergodic. Given the mathematical result of Markus and Meyer, where they argued that 'in the space of Hamiltonian dynamical systems, most Hamiltonians are neither ergodic nor integrable', there must be very few ergodic and integrable systems. This is supported by the fact that the only known ergodic systems are the billiard ball problems \cite{Sinai} and maybe the one-dimensional gravitational gas for certain number of particles (see for example the discussion in \cite{Yawn}, which discusses the contradicting results by various groups). As for complete integrability, which seems to have more known examples, what can be done is to look for a canonical transformation to ignorable coordinates. But the existence of such a transformation is a case to case basis, i.e., there is no general method for determining the existence of such a transformation. 
	
	The above discussions present what are rigorously established about systems that attain equilibrium. And they seem to be rather limited. But in practice though, thermodynamic systems do attain thermal equilibrium after a sufficiently long time. How do we explain these systems then? This is where the concept of $\epsilon$-ergodicity comes in, see  \cite{Frigg} for a clear discussion and references to the original literature. Thermodynamic systems must be ergodic in the phase space defined by the energy surface except for a small region of measure $\epsilon$. And because they are still mostly ergodic, these systems attain thermal equilibrium.
	
	These ideas that we summarized are still not settled and we will not deal with them in this paper. Our concern is the Liouville equation and its solution. The ansatze we present regarding its solution assumes attainment of equilibrium. We will show that for systems that attain thermal equilibrium, the ansatze has a simple solution. For the case of completely integrable systems that equilibrate with a generalized Gibbs ensemble, the method is no longer valid and it is best to solve the Liouville equation not in the original phase space but in the canonically transformed space with ignorable coordinates.      	        
\section{\label{sec:level2}The Solution Ansatze for Ergodic Systems} 
	We will consider a system of N particles as defined in Section I. The particle dynamics is given by a background potential $V(x_i)$ that acts on each particle and a 2 body potential $U(x_i-x_j)$. The Hamiltonian is given 
\begin{equation}\label{sha5}
H=\sum_{i=1}^N\left\{\frac{1}{2}\frac{p_i^2}{m_i}+\left[V(x_i)+\frac{1}{2}\sum_{j\neq i}U(x_i-x_j)\right]\right\}.
\end{equation}
The classical statistical mechanics of the system begins with the Gibbs distribution given by the Liouville equation
\begin{equation}\label{sha6}
\dfrac{\partial f_N}{\partial t}+\sum_{i=1}^N\left\{\dfrac{\partial H}{\partial p_i}\dfrac{\partial f_N}{\partial x_i}-\dfrac{\partial H}{\partial x_i}\dfrac{\partial f_N}{\partial p_i}\right\}=0.
\end{equation} 
We will assume a solution of the form
\begin{equation}\label{sha7}
f_N(x,p;t)=\int_0^\infty d\omega A(\omega) e^{-\omega t}F(x,p;\omega),
\end{equation}
where $A(\omega)$ is the weight of contribution of the particular mode $F(x,p;\omega)$ and as it will be shown below, this will have to be solved from the initial distribution $f_N(x,p;t=0)$. The $\omega$ integral is limited to positive values because the problems we will address are ergodic (achieves thermal equilibrium).  

	Substituting into the Liouville equation, we find
\begin{subequations}\label{sha8}
\begin{gather}
\sum_{i=0}^N\left\{-\frac{p_i}{m_i}\dfrac{\partial F}{\partial x_i}+\Gamma_i(x)\dfrac{\partial F}{p_i}\right\}=-\omega F,\label{first}\\
\Gamma_i(x)=\dfrac{dV(x_i)}{dx_i}+\frac{1}{2}\sum_{j\neq i}\dfrac{dU(x_i-x_j)}{dx_i}. 
\end{gather}
\end{subequations}
The crux of the paper is providing the solution to equations (7) and (8).

	The $\omega=0$ mode is also the equilibrium solution, i.e., $\dfrac{\partial f}{\partial t} =0$, as $t \rightarrow \infty$. We will separate this from $\omega \neq 0$ and write
\begin{equation}\label{sha9}
f_N(x,p;t)=A(0)F(x,p;0)+\int_\epsilon^\infty d\omega A(\omega) e^{-\omega t}F(x,p;\omega).
\end{equation} 
In this form, the the equilibrium distribution is clearly isolated. 

	We assume $F(x,p;0)=X(x)P(p)$, i.e., a separable solution. Substituting this ansatz in equation (8) with $\omega=0$, we find
\begin{equation}\label{sha10}
\sum_{i=0}^N\left\{-\frac{p_i}{m_i}\dfrac{d\ln X}{dx_i}+\Gamma_i(x)\dfrac{d\ln P}{dp_i}\right\}=0.
\end{equation}
It is obvious that equation (10) is solved by
\begin{subequations}\label{sha11}
\begin{gather}
\dfrac{d\ln X}{dx_i}=-\beta \Gamma_i(x),\label{first}\\
\dfrac{d\ln P}{dp_i}=-\beta\frac{p_i}{m_i},
\end{gather}
\end{subequations}
where the constant $\beta$ is identified with $\frac{1}{kT}$ and is introduced because of Gibbs idea of eventual thermal equilibrium for the system. The $\omega=0$ solution to equation (9) has the form of Maxwell-Boltzmann distribution.
\begin{equation}\label{sha12}
F(x,p;0)= F_{MB}=C\exp{(-\beta H)},
\end{equation}
where H is given by equation (5). The constant C will normalize the distribution.

	From hereon, the distribution we are solving in equation (9) should be labeled as $f_E(x,p,t)$ to emphasize the ergodic nature of the dynamics since the equilibrium distribution is the thermal distribution. 
	
	It is clear that the $\omega\neq0$ solution to equation (8) will not be separable. Let us rewrite this equation as
\begin{equation}\label{sha13}
\sum_{i=1}^N\left\{\frac{\partial}{\partial x_i}\left[-\frac{p_i}{m_i}\ln F\right]+\frac{\partial}{\partial p_i}\left[\Gamma_i(x)\ln F\right]\right\}=-\omega.
\end{equation}
The solution to this equation can be solved from
\begin{subequations}\label{sha14}
\begin{gather}
-\frac{p_i}{m_i}\ln F=-\frac{\omega}{2N}ax_i+\alpha_i(x,p),\label{first}\\
\Gamma_i(x)\ln F=-\frac{\omega}{2N}bp_i+\beta_i(x,p),
\end{gather}
\end{subequations}
where $\alpha_i(x,p)$ and $\beta_i(x,p)$ are unspecified functions at this point and a and b are constants that satisfy $a+b=2$. Before we solve these equations, let us clarify first how these equations should be treated. Equation (13) is a single first order partial differential equation for F. Equations (14a) and (14b) give 2N equations. For these equations to satisfy equation (13), they must always be taken together, for each $i=1,...,N$. Otherwise, this will become 2N equations with $(2N-1)!$ constraints, clearly an overspecified problem and thus not having a solution. 
	
	$\alpha_i(x,p)$ and $\beta_i(x,p)$ must satisfy  
\begin{equation}\label{sha15}
\sum_{i=1}^N\left\{\dfrac{\partial \alpha_i}{\partial x_i}+\dfrac{\partial \beta_i}{\partial p_i}\right\}=0.
\end{equation}
Equation (15) hints that the $\alpha_i(x,p)$ and $\beta_i(x,p)$ are like components of a 'solenoidal vector field' in phase space because its divergence is zero.  But the components of this 'solenoidal vector field' are not really independent because equations (14a) and (14b)give a relationship between $\alpha_i(x,p)$ and $\beta_i(x,p)$, i.e.,   
\begin{equation}\label{sha16}
\beta_i= \dfrac{a\frac{\omega}{2N}x_i\Gamma_i(x)-\alpha_i(x,p)\Gamma_i(x)+b\frac{\omega}{2N}\frac{p_i^2}{m_i}}{\frac{p_i}{m_i}}.
\end{equation}
Using equations (15) and (16), we find the equation satisfied by $\alpha_i$
\begin{equation}\label{sha17}
\sum_{i=1}^N\left\{\dfrac{\partial\alpha_i}{\partial x_i}-\frac{m_i}{p_i}\Gamma_i(x)\dfrac{\partial\alpha_i}{\partial p_i}+\frac{m_i}{p_i^2}\Gamma_i(x)\alpha_i\right\}=\sum_{i=1}^N\left\{\frac{a\omega}{2N}\dfrac{m_ix_i}{p_i^2}\Gamma_i(x)-\frac{b\omega}{2N}\right\}.
\end{equation}
	
	There are two observations that we make right away regarding $\alpha_i$ and $\beta_i$. First, we can multiply both of them by $N^\zeta$ and equation (14) will still satisfy equation (13) and equation (15) is still valid. We are free to choose the value of $\zeta$ later so that it will satisfy a certain condition. Second, from equation (17), we notice that $\alpha_i \propto \frac{\omega}{N}$ and by equation (16), the same is true with $\beta_i$. This means we can define \begin{subequations}\label{sha18}
\begin{gather}
\alpha_i=\frac{\omega}{N}\tilde{\alpha}_i,\label{first}\\
\beta_i=\frac{\omega}{N}\tilde{\beta}_i.
\end{gather}
\end{subequations}
Taking these two into account and solving for F by multiplying the two terms in equations (14a) and (14b), we find
\begin{equation}\label{sha19}
F(x,p,\omega)=\exp\left\{i\omega N^{(\zeta-1)}\left[\dfrac{\sum_{i=1}^N\left\langle \frac{abx_ip_i}{N^{2\zeta}}-\frac{\left(ax_i\tilde{\beta_i}+bp_i\tilde{\alpha_i}\right)}{N^{(\zeta-1)}|}+\tilde{\alpha_i}\tilde{\beta_i}\right\rangle}{\sum_{i=1}^N \frac{p_i}{m_i}\Gamma_i}\right]^{\frac{1}{2}}\right\},
\end{equation}
where $\tilde{\alpha_i}$ and $\tilde{\beta_i}$ satisfy equations similar to equations (16) and (17) but without $\frac{\omega}{N}$. 
	
	It will now be explained why we solved for F by multiplying equations (14a) and (14b) and then summing over i which resulted in equation (18). This F is oscillatory in phase space, which for large number of particles and or choice of $\zeta$ can be made to oscillate very fast. This will be important in solving for the coefficients $A(\omega)$.
  
	Since the F given by equation (19) is rapidly oscillating, then
\begin{equation}\label{sha20}
\int d^Nx d^Np F(x,p,\omega)=0.
\end{equation}
To justify why rapid oscillation of the integrand will lead to a zero integral, consider the following simple integral
\begin{equation}\label{sha21}
s(\lambda)=\int_{-\infty}^\infty dx \exp{(i\lambda f(x))},
\end{equation}
where $f(x)$ is a real function. The stationary phase approximation and perturbation theory gives
\begin{equation}\label{sha22}
s(\lambda)\propto \exp(i\lambda f(x_0))\lambda^{-\frac{1}{2}}\left[1+O(\lambda^{-2})+O(\lambda^{-3})+...\right],
\end{equation}
where $x_0$ is a minimum of f. Equation (21) clearly shows that $s(\lambda) \rightarrow 0$ as $\lambda \rightarrow \infty $.

	The only thing left is to solve for $A(\omega)$. Since the Gibbs distribution at any time t is normalized and so is the Maxwell Boltzmann distribution, integrating equation (9) in phase space and using equation (20) yields 
\begin{equation}\label{sha23}
A(0)=1.
\end{equation}
Suppose the system starts from an initial state with distribution $f_N(x,p,0)$, multiplying equation (9) at $t=0$ with $F^*(x,p,\omega')$ gives
\begin{equation}\label{sha24}
\int\left[f_N(x,p;0)-F_{MB}(x,p)\right]F^\ast(x,p;\omega')d^Nxd^Np=\int_\epsilon^\infty d\omega A(\omega)\left[\int d^Nxd^Np F(x,p;\omega)F^\ast(x,p;\omega')\right].
\end{equation}
We now prove that 
\begin{equation}\label{sha25}
\int d^Nxd^Np F(x,p;\omega)F^\ast(x,p;\omega') = \delta(\omega-\omega').
\end{equation}
From equation (19), when $\omega=\omega'$, the phase space integral diverges. When $\omega\ \neq \omega'$, the integrand oscillates very rapidly and the integral washes out giving zero. These prove equation (25), which then yields
\begin{equation}\label{sha26}
A(\omega)=\int\left[f_N(x,p;0)-F_{MB}(x,p)\right]F^\ast(x,p;\omega)d^Nxd^Np.
\end{equation}

	In summary, equations (9), (12), (19) (23) and (26) solve the Liouville equation for systems that thermal equilibrate. Equations (9) and (19) show that the Gibbs distribution is dominated by the equilibrium Maxwell-Boltzmann distribution with non-equilibrium, oscillatory (in phase space) contributions that dissipate in time. 

	To convince the reader that the solution described may be the only viable thermally equilibrating solution, let us solve equation (13) in another way. We rewrite this equation as
\begin{equation}\label{sha27}
\sum_{i=1}^N\left\{-\frac{p_i}{m_i}\dfrac{\partial\ln F}{\partial x_i}+\Gamma_i(x)\dfrac{\partial\ln F}{\partial p_i}\right\}=-\omega.
\end{equation}
Equation (27) is solved by
\begin{subequations}\label{sha28}
\begin{gather}
\dfrac{\partial\ln F}{\partial x_i}=\Gamma_i+\frac{\omega}{2N}\frac{1}{p_i/m_i}+\dfrac{\partial D(x,p)}{\partial x_i},\label{first}\\
\dfrac{\partial\ln F}{\partial p_i}=p_i/m_i-\frac{\omega}{2N}\frac{1}{\Gamma_i(x)}+\dfrac{\partial E(x,p)}{\partial p_i},
\end{gather}
\end{subequations}
provided D(x,p) and E(x,p) satisfy
\begin{equation}\label{sha29}
\sum_{i=1}^N\left\{-\frac{p_i}{m_i}\dfrac{\partial D}{\partial x_i}+\Gamma_i(x)\dfrac{\partial E}{\partial p_i}\right\}=0.
\end{equation}
However, D and E are not independent as equations (28;a,b) show. From these equations, we find 
\begin{equation}\label{sha30}
E(x,p)=\sum_{i=1}^N\left\{\frac{\omega}{2N}\frac{x_im_i}{p_i}+\frac{\omega}{2N}\frac{p_i}{\Gamma_i(x)}-\frac{p_i^2}{2m_i}+\int \Gamma_i(x'_i,x)dx'_i\right\}+D(x,p)
\end{equation}
Using equation (30) in equation (29), we find that D must be solved from
\begin{equation}\label{sha31}
\left\{\sum_{i=1}^N\left[-\frac{p_i}{m_i}\dfrac{\partial}{\partial x_i}+\Gamma_i(x)\dfrac{\partial}{\partial p_i}\right]\right\}D(x,p)=\sum_{i=1}^N\left\{-\frac{\omega}{2N}+\frac{p_i}{m_i}\Gamma_i(x)+\frac{\omega}{2N}\frac{x_im_i}{p_i^2}\Gamma_i(x)\right\}.
\end{equation}
Unfortunately, the operator that acts on D(x,p) is a singular operator with many zero modes - any function of the Hamiltonian H of the system is a zero mode since H is itself a zero mode. For equation (31) to have a square-integrable solution, the zero mode must be orthogonal to the source of equation (31). It is easy to see that this will never happen. The inner product of H with the source of equation (31) will never be zero. Thus, equation (31) does not have a solution and equation (27) cannot be solved by equation (28;a,b). This is the reason why we stated that the solution to the Liouville equation defined in the earlier part of this section may just be the only possible solution that equilibrates to the Maxwell-Boltzmann distribution.  	 	

\section{\label{sec:level3}Gravitational Gas}	
	To illustrate the formalism discussed in the earlier part of the previous section, we will consider the case of the one-dimensional gas. The ergodicity of this system is apparently debatable (see for example \cite{Yawn}) but we will use it anyway because of solvability. The potentials for this system are $V=0$ and $U(x_i-x_j)=\lambda\left|x_i-x_j\right|$. The function $\Gamma_i(x)$ is given by
\begin{equation}\label{sha32}
\Gamma_i(x)=\frac{\lambda}{2}\sum_{j\neq i}\left\{\Theta(x_i-x_j)-\Theta(x_j-x_i)\right\},
\end{equation}
where $\Theta(x_i-x_j)$ is the step function (equal to 1 for $x_j > x_i$, zero otherwise).

	From equations (17) and (18), $\tilde{\alpha_i}$ satisfies
\begin{equation}\label{sha33}
\sum_{i=1}^N\left\{\dfrac{\partial\tilde{\alpha_i}}{\partial x_i}-\frac{m_i}{p_i}\Gamma_i(x)\dfrac{\partial\tilde{\alpha_i}}{\partial p_i}+\frac{m_i}{p_i^2}\Gamma_i(x)\tilde{\alpha_i}\right\}=\sum_{i=1}^N\left\{\frac{a}{2}\dfrac{m_ix_i}{p_i^2}\Gamma_i(x)-\frac{b}{2}\right\}.
\end{equation} 
The solution to this equation is
\begin{equation}\label{sha34}
\tilde{\alpha_i}=\frac{a}{2}x_i+\frac{p_i^2}{m_i}\left(\frac{1}{\Gamma_i(x)}\right).
\end{equation}
Substituting equation (34) in equation (33), we find a remainder proportional to
\begin{equation}\label{sha35}
\sum_{i=1}^N\dfrac{1}{\Gamma_i^2}\dfrac{\partial\Gamma_i}{\partial x_i}=\sum_{i=1}^N\dfrac{\sum_{j\neq i}\delta(x_i-x_j)}{\sum_{j,k\neq i}\left\{\left[\Theta(x_i-x_j)-\Theta(x_j-x_i)\right]\left[\Theta(x_i-x_k)-\Theta(x_k-x_i)\right]\right\}}
\end{equation}
Taking into account the property of the step function and the possible positions of $(x_i,x_j,x_k)$, we find that each i term has the denominator proportional to $4N^2$. Furthermore to the extent that we can neglect the momentary collission between elements of the gravitational gas, the huge denominator ($N^2$) makes the remainder negligible. Equation (34) then gives the solution to equation (33). 

	To complete the solution for F, we have to solve for $\tilde{\beta_i}$. From equations (16) and (18), 
\begin{equation}\label{sha36}
\tilde{\beta_i}= \dfrac{\frac{a}{2}x_i\Gamma_i(x)-\tilde{\alpha_i}(x,p)\Gamma_i(x)+\frac{b}{2}\frac{p_i^2}{m_i}}{\frac{p_i}{m_i}}.
\end{equation} 
Substituting equation (34), we find
\begin{equation}\label{sha37}
\tilde{\beta_i}=\left(\frac{b}{2}-1\right)p_i.
\end{equation}
We can now substitute equations (37) and (34) in equation (19) to find $F(x,p,\omega)$ and from there evaluate $A(\omega)$ using equation (26) if we know the initial Gibbs distribution. Thus, we have given the Gibbs distribution from the Liouville equation of a classical 1D gravitational gas. 

\section{\label{sec:level4}Completely Integrable Systems} 

	In this section, we will address how to solve the Liouville equation for completely integrable systems. But before we present the method that specifically applies to these systems, we first argue why the method we used for systems that attain thermal equilibrium does not hold in this case. Since completely integrable systems also attain equilibrium, albeit with a generalized Gibbs ensemble, why can we not use the ansatze given by equation (9) with an equilibrium distribution given by equation (3) instead of equation (12)? There is one simple reason. Since completely integrable systems have N integral invariants $I_i(x,p)$ that are in involution (satisfies equation (4)), the system is essentially a constrained system but not in the way of Dirac's constrained system where the constraints arise from solving the velocities from the momenta and carrying out a consistency iteration \cite{Dirac}. The Poisson brackets must be modified because the system is constrained in the N tori and the geometry of these N tori must result in different brackets between the coordinates and momenta. 
	
	How then are the Poisson brackets changed? The solution begins from the fact that canonical transformations form a group. And in the case of completely integrable systems, the integral invariants are the generators of the infinitesimal canonical transformations. Furthermore, these integral invariants commute because they are in involution, thus hinting of an Abelian group structure in phase space. In the language of Dirac formalism, the integral invariants form a set of first-class constraints. To derive the modified Poisson brackets, these first-class constraints need to be made second-class and this is done by imposing subsidiary conditions, which in the language of gauge theory are known as gauge-fixing conditions. There is one gauge condition, say $G_i$, for every first-class constraint and they are generally made to satisfy
\begin{subequations}\label{sha38}
\begin{gather}
\left\{G_i,G_j\right\}=0,\label{first}\\
\left\{G_i,I_j\right\}\neq0.
\end{gather}
\end{subequations} 
These equations will guarantee that the subsidiary conditions are independent of the integral invariants. Now that the integral invariants are now made second-class, the Poisson brackets can be modified following Dirac's prescription.    
	
	To make these discussions more explicit, consider a canonical transformation from $(x_i,p_i)$ to $(Q_i,P_i)$ via a generating function $W(x_i,Q_i,t)$ such that the new Hamiltonian
\begin{equation}\label{sha39}
K=H+\dfrac{\partial W}{\partial t},
\end{equation}  
is independent of all the $Q_i$, with $i=1,...N$, i.e., all the new coordinates are ignorable \cite{Goldstein}. Since the equations of motion given by equation (1) remain the same in the new coordinates and momenta using the new Hamiltonian K, we find that
\begin{equation}\label{sha40} 
\dot{P}_i=0.
\end{equation}
We will make the further assumption that the generating function W is time independent so that the new coordinates and momenta are only functions of the old coordinates and momenta and not explicitly time dependent. Equation (40) means $P_i(x,p)=P_{0i}$, i.e., all the momenta are constants of motion given by the initial values, leading to the integral invariants $I_i=P_i(x,p)-P_{0i}$. Since the fundamental Poisson brackets are invariant under canonical transformation, i.e., $\left\{p_i,p_j\right\}=\left\{P_i,P_j\right\}=0$, then the integral invariants $I_i$ satisfy equation (2), they are in involution. The infinitesimal canonical transformations are now given by
\begin{subequations}\label{sha41}
\begin{gather}
\delta x_i=\epsilon_j\left\{x_i,I_j\right\}=\epsilon_j\dfrac{\partial P_j}{\partial p_i},\label{first}\\
\delta p_i=\epsilon_j\left\{p_i,I_j\right\}=-\epsilon_j\dfrac{\partial P_j}{\partial x_i}
\end{gather}
\end{subequations}
where j is summed and $\epsilon_j$ are the infinitesimal parameters of the canonical transformation. Equation (41) presents the active view of a canonical transformation and hints of a 'gauge' structure. Once we choose a 'gauge' condition $G_i=0$, the combined conditions $(I_i,G_i)=(C_a)$, with a = 1,..., 2N will now be second-class. All Poisson brackets, including the fundamental ones involving $(x_i,p_i)$ and those that appear in the Liouville equation must now be replaced by Dirac brackets given by
\begin{equation}\label{sha42}
\left\{A,B\right\}_{DB}=\left\{A,B\right\}-\left\{A,C_a\right\}\left\{C_a,C_b\right\}^{-1}\left\{C_b,B\right\}.
\end{equation}  
Note that because of equations (2) and (38), the matrix $\left\{C_a,C_b\right\}$ is off-diagonal and the inverse is guaranteed to exist. This is the reason why we can no longer make use of the operations in Section II to derive the full-time dependent Liouville distribution for completely integrable systems, with the GGE as the equilibrium distribution, in the (x,p) phase space. 

	Before we solve the Liouville equation, let us discuss the context of the gauge theory connection observed above. The link between gauge theories and integrable systems had been explored by mathematicians for more than twenty years now, see for example \cite{Olshanetsky1} for a review of the literature. The link with gauge theories followed the connection of completely integrable systems with semi-simple Lie algebras, which was established a decade earlier \cite{Olshanetsky2}. The mathematical discussion of the connection makes use of the language of differential geometry in symplectic space and Lax representation. 
	
	In this paper, we use the language and methods of gauge theory that physicists are more familiar with. The N integral invariants $I_i$ that are in involution, i.e., commuting, suggests an abelian gauge structure with N $U(1)$. However, the integral invariants can also satisfy a non-Abelian algebra, which will make the Dirac brackets a bit more involved. The conserved quantities are the integral invariants themselves and as seen in equation (41), they are also the generators of the gauge transformation, i.e., the canonical transformation that leave Hamilton's equations invariant. Following the treatment of Dirac for systems with constraints, the integral invariants are therefore all first-class constraints. Thus, we need to make them second-class by imposing gauge-fixing conditions $G_i$. This in turn means replacing the Poisson brackets in the Liouville equation by Dirac brackets as given in equation (42). The distribution is now given by
\begin{equation}\label{sha43}
\dfrac{\partial f}{\partial t}=\left\{H,f\right\}-\sum_{i,j}^N\left\{H,G_i\right\}\left\{G_i,I_j\right\}^{-1}\left\{I_j,f\right\}.
\end{equation}
The derivation of this equation also follows the derivation of equation (6), only $\dot{x}_i$ and $\dot{p}_i$ are given by Dirac brackets to take into account the integral invariants, that essentially act as first-class constraints. Since equation (43) is dependent on a choice of gauge given by the $G_i$, it gives the impression that the solution to the Liouville equation is dependent on the choice of gauge. This will not make sense because the distribution is a physical quantity, all thermodynamic quantities are calculated from it via averaging. For equation (43) then to make sense, its solution must be independent of $G_i$. This is indeed true as the following arguments show. Let the solution of equation (43) be $f(x,p,t;G)$ which must be normalized, i.e.,
\begin{equation}\label{sha44}
\int dx_idp_i f(x,p,t;G)=1.
\end{equation}
Taking the functional derivative of f with $G_i$, we find that its phase space integral is zero. This does not guarantee that $\dfrac{\delta f}{\delta G_i}=0$. This only follows if this quantity is also a distribution, i.e., it is positive semi-definite. Functionally differentiating equation (43) with $G_i$, we see that $\dfrac{\delta f}{\delta G_i}$ also satisfies equation (43) just like the distribution f. Thus it is a also a distribution with zero norm as shown by equation (44). Thus $\dfrac{\delta f}{\delta G_i}$ must be zero. This proves that the solution of equation (43) is independent of the choice of gauge.      	
	
	Instead of solving equation (43), we make use of a more direct approach, which is analogous to choosing a unitarity gauge in gauge theories with spontaneous symmetry breaking (again, the language of physicists). The reason for this comparison is that in the unitarity gauge in these theories, the physical degrees of freedom (for example in electro-weak theory, massive W and Z bosons, massless photons and the presence of a Higgs scalar) are explicit. In completely integrable systems, choosing the coordinates and momenta $(Q_i,P_i)$ such that all $Q_i$ are ignorable leads to a very simple description - all momenta are constants and all the coordinates $Q_i$ evolve linearly in time and the phase space is simply given by equation (51), i.e., the system's dynamics is rather simple. Thus, solving a completely integrable system with ignorable coordinates may be compared to a unitarity gauge description in gauge theories.  

	Recall first the equilibrium distribution, as derived using Jaynes' method, is given by    
\begin{equation}\label{sha45}
f_{GGE}=Z^{-1}(a_i)\exp{(-\sum_{i=1}^Na_iP_i(x,p))}.
\end{equation} 
In this section, we will provide another derivation of this distribution making use of the canonical transformation properties of the distributions and solving the distribution in the phase space with ignorable coordinates, i.e., in what we labeled as the unitarity gauge. We will find out that the distribution in this phase space is time-independent. Then, we show that we can express this in the form of a generalized Gibbs ensemble and derive an expression for the coefficients in terms of initial conditions. 

	Consider first a general canonical transformation from (x,p) to (Q,P). We call the distribution in (Q,P) $\tilde{f}$. Since the distributions are normalized, i.e.,
\begin{equation}\label{sha46}
\int d^Nxd^Np f(x,p,t)=\int d^NQd^NP \tilde{f}(Q,P,t)=1,
\end{equation}
and the Jacobian of the canonical transformation from (x,p) to (Q,P) is $\pm$1 \cite{Goldstein}, then $\tilde{f}=\pm f$. We take the positive value because we are dealing with distributions. 

	This also follows from the Liouville equation. Starting from the Liouville equation for $\tilde{f}$ given by
\begin{equation}\label{sha47}
\dfrac{\partial \tilde{f}}{\partial t}+\sum_{i=1}^N\left(\dfrac{\partial \tilde{f}}{\partial Q_i}\dot{Q}_i+\dfrac{\partial \tilde{f}}{\partial P_i}\dot{P}_i\right)=0.
\end{equation}	
Using $Q_i=Q_i(x,p)$ and $P_i=P_i(x,p)$, the above equation is exactly the same equation satisfied by f(x,p,t). Thus, $\tilde{f}=f$ under a canonical transformation.

	If we now make the canonical transformation from (x,p) to (Q,P) satisfy the completely integrable condition as discussed at the top of this section, we will label the distributions as $f_{CI}(x,p,t)$ and $\tilde{f}_{CI}(Q,P,t)$. We will solve the distribution in the (Q,P) phase space and taking note of the fact that $\dot{P}_i=0$, the distribution satisfies
\begin{equation}\label{sha48}
\dfrac{\partial \tilde{f}_{CI}}{\partial t}+\sum_{i=1}^N\dfrac{\partial \tilde{f}_{CI}}{\partial Q_i}\dot{Q}_i. 	
\end{equation}	
But the equation of motion for the coordinates $Q_i$ reads
\begin{equation}\label{sha49}
\dot{Q}_i=\dfrac{\partial K}{\partial P_i}=\omega_i,
\end{equation}
where the $\omega_i$ are constants dependent on the constant momenta $P_{0i}$ since the Hamiltonian K is purely P dependent. Integrating this equation gives
\begin{equation}\label{sha50}
Q_i=\omega_i t+\lambda_i, 
\end{equation}
where $\lambda_i$ are constants that depend on the initial state of the system. The classical dynamics of the integrable system expressed in terms of (Q,P) is much simpler. In this phase space though, the system only evolves in the region defined by
\begin{subequations}\label{sha51}
\begin{gather}
P_i=P_{0i},\label{first}\\
\dfrac{Q_1-\lambda_1}{\omega_1}=\dfrac{Q_2-\lambda_2}{\omega_2}=....=\dfrac{Q_N-\lambda_N}{\omega_N}.
\end{gather}
\end{subequations} 
This means the system is fixed in the momentum space and evolves only in the coordinate space defined by (51b).

	Substituting equation (49) in equation (48), we find that the distribution $\tilde{f}_{CI}$ must be only a function of t and the coordinates $Q_i$. Equation(48) can be solved by separation of variables, i.e.,
\begin{equation}\label{sha52}
\tilde{f}_{CI}=T(t)G(Q)
\end{equation}
Surprisingly, the solution is time-independent and is given by
\begin{equation}\label{sha53}
\tilde{f}_{CI}=\exp{\left(-\sum_{i=1}^Nb_i\frac{\lambda_i}{\omega_i}\right)},
\end{equation}
where the constants $b_i$ is normalized to 1, i.e., $\sum_{i=1}^Nb_i=1$. Since the terms in the above distribution are all constants, the distribution of the completely integrable system in the phase space with ignorable coordinates can be expressed this in terms of another set of constants, the momenta $P_{0i}$, 
\begin{subequations}\label{sha54}
\begin{gather}
\tilde{f}_{CI}=\exp{\left(-\sum_{i=1}^Na_iP_{0i}\right)},\label{first}\\
a_i=b_i\frac{\lambda_i}{\omega_iP_{0i}},
\end{gather}
\end{subequations}
and in this form, it is the same as the generalized Gibbs ensemble with the Lagrange multiplier given in terms of the initial conditions.
   
	Using the transformation of distributions under a canonical transformation and equivalently using the arguments for gauge-invariance, we find that in the original phase space (x,p), the distribution is time-independent and given by 
\begin{equation}\label{sha55}
f_{CI}(x,p,t)=\tilde{f}_{CI}=\exp{\left(-\sum_{i=1}^Na_iP_i(x,p)\right)}.
\end{equation}
Thus, the distribution of a completely integrable system at any time is given by the generalized Gibbs ensemble.
\section{\label{sec:level5}Two Particles Interacting via Central Potential}
	We will illustrate the method discussed in the previous section on a problem that is completely integrable. Two particles in 3D interacting via a central potential is known to be a completely integrable system, see for example \cite{Masoliver}. The integral invariants are - the Hamiltonian, the center of mass momentum, the square of the total angular momentum and the z component of the total angular momentum - for a total of 6, which equals the number of degrees of freedom. The method in the previous section however does not apply completely because for this example, not all the coordinates can be made ignorable. The distance between the two particles cannot be made ignorable.
	
	The dynamics is defined by the Hamiltonian
\begin{equation}\label{sha56}
H=\dfrac{\vec{p_1}^2}{2m_1}+\dfrac{\vec{p_2}^2}{2m_2}+V(\left|\vec{r_2}-\vec{r_1}\right|).
\end{equation}
Transforming to center of mass and relative coordinates and momentum given by
\begin{subequations}\label{sha57}
\begin{gather}
\vec{R}=\dfrac{m_1\vec{r_1}+m_2\vec{r_2}}{m_1+m_2},\label{first}\\
\vec{r}=\vec{r_2}-\vec{r_1},\label{second}\\
\vec{P}=M\dot{\vec{R}}=\vec{p_1}+\vec{p_2},\label{third}\\
\vec{p}=\mu\dot{\vec{r}}=\frac{m_1}{M}\vec{p_2}-\frac{m_2}{M}\vec{p_1},
\end{gather}
\end{subequations}
where $M=m_1+m_2$ is the total mass and $\mu=\frac{m_1m_2}{m_1+m_2}$ is the reduced mass. In the new coordinates and momentum, the Hamiltonian becomes
\begin{equation}\label{sha58}
H=\dfrac{\vec{P}^2}{2M}+\dfrac{\vec{p}^2}{2\mu}+V(r),
\end{equation}
which clearly shows that the center of mass coordinate $\vec{R}$ is ignorable. 
	Since H is separable into the center of mass and relative coordinates and momenta and $\vec{R}$ is ignorable, the Gibbs distribution can be written as
\begin{equation}\label{sha59}
f(X_i,P_i,x_i,p_i,t)=f_{CM}(X_i,P_i,t)f_{rel}(x_i,p_i,t),
\end{equation}
where $f_{CM}$ and $f_{rel}$ satisfy their respective Liouville equations
\begin{subequations}\label{sha60}
\begin{gather}
\dfrac{\partial f_{CM}}{\partial t}=-\sum_{i=1}^3 \frac{P_i}{M}\dfrac{\partial f_{CM}}{\partial X_i},\label{first}\\
\dfrac{\partial f_{rel}}{\partial t}=\sum_{i=1}^3\left[\dfrac{\partial V}{\partial x_i}\dfrac{\partial f_{rel}}{\partial p_i}-\frac{p_i}{\mu}\dfrac{\partial f_{rel}}{\partial x_i}\right].
\end{gather}
\end{subequations}
The center of mass Liouville equation can be solved by separation of variables into a function of t and a function of $X_i$ and the answer is
\begin{equation}\label{sha61}
f_{CM}=\textsl{A}\exp{\left[\sum_{i=1}^3c_i(X_i-\frac{P_i}{M}t)\right]},
\end{equation}
where the $c_i$ are constants. If we consider the classical equations of motion,
\begin{subequations}\label{sha62}
\begin{gather}
P_i = P_{0i},\label{first}\\
X_i=\frac{P_{0i}}{M}t+X_{0i},
\end{gather}
\end{subequations}
then the Gibbs distribution becomes
\begin{equation}\label{sha63}
f_{CM}=\textsl{A}\exp{\left[\sum_{i=1}^3c_iX_{0i}\right]},
\end{equation}
a constant, a distribution of the same form as equation (54).
	
	The distribution for the relative coordinates is more involved, but follows the same procedure. First, we rewrite the Hamiltonian in spherical coordinates
\begin{equation}\label{sha64}
H=\dfrac{p_r^2}{2\mu}+\dfrac{p_\theta^2}{2\mu r^2}+\dfrac{p_\phi^2}{2\mu r^2 \sin^2\theta}+V(r),
\end{equation}
which shows that $\phi$ is ignorable. The Liouville equation with this Hamiltonian is not easy to solve. We simplify the Hamiltonian by taking into account that the dynamics is essentially a 2D problem, which we can assume to lie on the (x,y) plane (or $\theta=0$). The angular momentum is $p_\phi=\mu r^2 \dot{\phi}$ is conserved, which gives another integral invariant. The Hamiltonian is
\begin{equation}\label{sha65}
H=\dfrac{p_r^2}{2\mu}+\dfrac{p_\phi^2}{2\mu r^2}+V(r).
\end{equation} 
The Liouville equation becomes
\begin{equation}\label{sha66}
\dfrac{\partial f_{rel}}{\partial t}=-\frac{p_r}{\mu}\dfrac{\partial f_{rel}}{\partial r}+\left(\dfrac{dV}{dr}-\frac{p_\phi^2}{\mu r^3}\right)\dfrac{\partial f_{rel}}{\partial p_r}-\frac{p_\phi}{\mu r^2}\dfrac{\partial f_{rel}}{\partial \phi}.
\end{equation}
This equation is still difficult to solve so we make use of the energy integral invariant. For a given energy E, which is conserved, the momentum $p_r$ can be solved from equation (65), which then simplifies the Liouville equation to
\begin{equation}\label{sha67}
\dfrac{\partial f_{rel}}{\partial t}=-2\left\{\frac{2}{\mu}\left[E-V(r)-\frac{p_\phi^2}{2\mu r^2}\right]\right\}^{\frac{1}{2}}\dfrac{\partial f_{rel}}{\partial r}-\frac{p_\phi}{\mu r^2}\dfrac{\partial f_{rel}}{\partial \phi}.
\end{equation}
This equation is solved by separation of variables
\begin{equation}\label{sha68}
f_{rel}=T(t)\Psi(r,\phi).
\end{equation}
The time part is easy, 
\begin{equation}\label{sha69}
T(t)=T_0\exp{(-et)}. 
\end{equation}
The $(r,\phi)$ is solved from
\begin{equation}\label{sha70}
2\mu r^2\textit{S}(r)\dfrac{\partial \Psi}{\partial r}+p_\phi\dfrac{\partial \Psi}{\partial \phi}=e\mu r^2\Psi,
\end{equation}
where
\begin{equation}\label{sha71}
\textit{S}(r)=\int_{r_0}^r ds\left\{\frac{2}{\mu}\left[E-V(s)-\frac{p_\phi^2}{2\mu s^2}\right]\right\}^{-\frac{1}{2}}.
\end{equation}
The solution to this equation, taking into account the $2\pi$ periodicity in $\phi$ is 
\begin{equation}\label{sha72}
\Psi(r,\phi)=\psi_0\exp{\left\{\int_{r_0}^r\left[\frac{e}{2\textit{S}(r')}\mp\dfrac{inp_\phi}{2\mu r'^2\textit{S}(r')}\right]dr'\pm in\phi \right\}}.
\end{equation}
The distribution $f_{rel}$ is given by multiplying together equations ((69) and (72). 

	At this stage, we have not taken into account the solutions to the equations of motion, which are
\begin{subequations}\label{sha73}
\begin{gather}
t=\textit{S}(r),\label{first}\\
\phi=\phi_0+\frac{p_\phi}{\mu}\int_0^t \dfrac{dt'}{r^2(t')}.
\end{gather}
\end{subequations}
Using equation (73a), we can express the $f_{rel}$ independent of time and function of $(r,\phi)$. If we use both (73a) and (73b), the distribution is a complicated function of r alone.

	Finally, the complete distribution can be found by multiplying together $f_{CM}$ and $f_{rel}$.  This completes the solution to the Liouville equation for this particular completely integrable system. 		 	
\section{\label{sec:level6}Conclusion} 
	In this paper, we showed the solution to the Liouville equation for systems that attain thermal equilibrium and completely integrable systems. For systems that achieve thermal equilibrium, we used an ansatze that clearly shows the Maxwell-Boltzmann distribution at equilibrium. We showed what may be the only possible direct solution of the Liouville equation. For completely integrable systems, the group properties of the integral invariants allowed us to use the formalism of Dirac's constrained formalism to show that the Liouville equation in the original phase space is rather complicated but independent of the choice of gauge ($G_i$). Instead, we transformed to the phase space with ignorable coordinates because the system's dynamics is rather simple (constant momenta and coordinates evolving linearly in time) and thus we compare this to the unitarity gauge description of gauge theories. The Liouville equation is also easily solved resulting in the generalized Gibbs ensemble distribution. 
	
	There are two problems that arise from this paper. The first is to show that equation (54b) gives the same result as equation (4). In the literature today, equation (4) has only been verified (in quantum systems) in a few cases \cite{Calabrese}, \cite{Cassidi}. A proof of the equality of (54b) and (4) is needed. The second is what is the BRST symmetry of completely integrable systems? This follows from the fact that the integral invariants that are in involution are essentially first-class, Abelian constraints. We know from gauge theories that after fixing the gauge there is a leftover global symmetry involving fermionic degrees of freedom, the BRST symmetry, which is important in proving the renormalizability of gauge theories. For completely integrable systems there must also be a BRST symmetry and this is what we show in the next paper, which we also apply to the example discussed in section 5 \cite{Magpantay}.   	
	
	As a final comment, we ask what happens in the quantum case. For completely integrable systems, the literature shows many papers dealing with the quantum generalized Gibbs ensemble. For ergodic systems, a quantum treatment of the Sinai billiard problem had been done by Berry \cite{Berry}. To extend section II to the quantum case, we can proceed by solving the quantum Liouville equation. But since the coordinates and momenta no longer commute, the mathematical manipulations in the second section are no longer valid because products of functions must be replaced by the Groenewold-Moyal product \cite{Groenewold}, \cite{Moyal}. Solving the quantum Liouville equation is a lot more complicated than the classical Liouville equation. There is another way to incorporate quantum corrections to classical statistical mechanics and that is through the Wigner distribution \cite{Wigner}, a starting point in the phase space formulation of quantum mechanics. A perturbative solution to the Wigner equation is easily written down because the $\hbar^0$ term is just the Liouville equation, the first quantum correction is order $\hbar^2$ and the next is $\hbar^4$, etc. For all intents and purposes, the first quantum correction will sufffice. Since we already have a solution to the classical Liouville equation as defined in the section II, we can just extend this to the quantum case by solving the $\hbar^2$ correction to the Gibbs distribution. This is now being worked out for a non-trivial system (the third derivative of the potential must be non-zero).     

\begin{acknowledgments}
JAM would like to thank the Creative Work and Research Grant of the University of the Philippines System for supporting his research. He also would like to thank Perry Esguerra for discussions. CUMP would also like to thank Pecier Decierdo for discussions. We also acknowledge communications on the first draft of the paper that led to the changes made in this second version.
\end{acknowledgments}

\end{document}